# The ROSAT HRI Point Spread Function

**Peter Predehl** [1] **and Almudena Prieto** [2]

1 Max-Planck-Institut für extraterrestrische Physik, Giessenbachstrasse, D-85748 Garching, email: predehl@mpe.mpg.de

2 European Southern Observatory, Karl-Schwarzschildstrasse 2, D-85748 Garching, email: aprieto@eso.org

## Abstract

A sample of the brightest point-like sources observed with the ROSAT-HRI were analysed to asses on the intrinsic shape of the ROSAT-HRI Point Spread Function (PSF). Almost all of the HRI observations collected during the ROSAT lifetime are found to be artificially broadened by factors up two ~2 due to residual errors in the ROSAT aspect solution. After correction by departing pointing positions, the width of the core of the PSF is found to be less than 5 arcsec (half energy width, HEW). On the basis of these results, an improved analytical representation of the ROSAT-HRI PSF is provided. However, for most of the new observations the source countrate is too weak to allow reliable recovering pf the ROSAT-HRI resolution. Therefore, a series of examples (data, correction, and theoretical PSF) are given in order to help the user in determining whether "his source" is extended or not.

## The ROSAT Attitude Measurement and Control System

The ROSAT Attitude Measurement and Control System (AMCS) had a series of problems since the beginning of the mission. The AMCS contained several components for the measurement of the attitude:

- a sun-sensor measured the coarse orientation of the solar panels towards the sun. This system served primarily for maintaining the health of the instrument but not for the actual measurement of the telescope's attitude. Loosing the sun-orientation led to a safe-mode of the entire spacecraft leading to a switch off of all science instruments.
- 4 gyroscopes (out of which three were needed always) delivered information about the slew-speed and were used for the reorientation of the satellite mainly.
- ROSAT had two startrackers not only for redundancy reasons. Their field of views were slightly displaced with respect to each other in order to enhance the chance of finding enough guide stars (three were needed). A catalogue of guide stars within the actual great circle on the sky (±15° perpendicular to the sun-direction) was stored in the attitude computer.
- a magnetometer was originally planned only for the de-saturation of the reaction wheels via magnetic coils (see below).

The control part of the AMCS comprised four reaction wheels (out of which three were absolutely needed). The AMCS suffered from a whole series of failures during the mission:

- after about three months of operation the first star tracker failed for unknown

reasons; however, the mission could still be performed without any loss of performance.

- within the first year of the mission two out of the four gyroscopes failed, a third one lost part of its performance later. This degradation was compensated by a major change of the onboard "intelligence"': instead of relying on the gyroscopes during a slew, the information of the sun-sensor and the magnetometer was redirected into the AMCS software. Thus, the mission could still be sustained without degradation but occasionely ROSAT felt into safemode whenever guide stars were lost completely.
- after eight years, the second star tracker also failed. Now we tried to use the information of the auxiliary startracker which belonged to a separate instrument, the XUV Wide Field Camera (WFC). Various attempts failed more or less. Finally, the telescope faced towards the sun accidently. This led to the loss of the HRI. Only a few observations with the PSPC (which still had a small amount of gas available) could be made. A few weeks later, at the end of 1998, ROSAT was switched off finally.

## ROSAT HRI error budget

Prior to launch, the total error budget regarding the telescope's performance was verified by tests and analytical studies. Several contributions lead to both blur and positional inaccuracy of (point-) source images (Hefler, 1989). It turned out during the development of the ROSAT telescope that the mirror system performed much better than specified. Therefore we had to introduce the "wobble-mode", a slight and periodic reorientation of the pointing direction in order to avoid that point sources were obstructed by the support-grid structures of the detector entrance windows. The complete error budget as specified and measured is given in Table 1.

Table 1: Error Budget of the ROSAT-HRI Attitude accuracy

| Error Terms | Specification | Actual Values |
|---|---|---|
| A) Line of Sight (LOS) | | |
| Attitude Measurement and Control System | 3.0" | 2.6" |
| Boresight | 4.5" | 2.8" |
| Fiducial Light System | 2.5" | |
| sum LOS | | 3.82" |
| B) Blur Circle (BC) | | |
| Mirror Assembly | 3.5" | 1.8" |
| Mirror Assembly Thermal Control | 1.0" | 0.86" |
| Structure Telescope | 4.0" | 0.6" |
| Focal Instrumentation | 3.0" | 1.33" |
| sum BC | | 2.49" |
| Total Sum | 10" | 6.4" |

## In-orbit performance of the AMCS

At the begin of the mission, while the boresight was not yet performed, blurring of sources occurred, primarily during the scan-phase. During the pointed phase of the mission, a slight blurring due to residuals of the wobble-correction could not be avoided. Furthermore, a slight displacement between different OBIs (observation intervals) of the same object was also unavoidable, probably due to a boresight-error. Altogether, most of the HRI observations showed a spatial smearing of the order of less than 10 arcsec, less than the specification but clearly to be seen because the actual performance of the telescope could do better. The intrinsic spatial resolution of the PSPC (about 25 arcsec) was too low for this smearing to really contribute to a degradation of PSPC observations. In particular with the HRI, residual errors in the aspect solution lead to artificially elongated and broadened images by factors up to ~2. All HRI data is affected, however, the quantification of the effect, i.e, the broadening of the PSF, was difficult to detect, particularly for faint sources while, for brighter sources, this effect could be corrected (see below).

**The Point Response Function (PSF)**

The results derived above have been used to derive an improve analytical representation of the HRI PSF. The ROSAT HRI PSF consists of the following terms:

1. Core width, almost independent of energy. Fixed to the value found above.
2. Mirror scattering: depends on energy. This dependence has accurately been measured using the PSPC.
3. Off-axis dependent broadening

An analytical representation of the standard PSF can be found in David et al (1999) or Boese (2000), the computer codes (IDL) for a calculation of the standard PSF and the modified PSF are given in the annex of this report.

**Recovering the intrinsic HRI PSF**

The degradation of the PSF with respect to the nominal telescope performance is caused by an image-shift which varies in time. The recovery of the HRI-PSF is very similar to the tip/tilt-mechanism used in optical telescopes: the centroids of a point source (or an X-ray source serving as "guide star", respectively) is determined within short time intervals. The images of all intervals are re-centered giving a corrected complete image. This procedure has to be optimized with regard to photon statistics, length of intervals and the size of the area in which the centroids are calculated. It is easily to realize that the methods works satisfactorily only if the photon statistics allow to make the time intervals short compared with the wobble period or other time scales of image-shifts. Morse (1994) has used this method based on photons in detector coordinates while we prefer to do it directly in celestial coordinates. Harris et al. (1998) suggested to take intervals not in time but in phase of the wobble period in order to increase the statistical significance. This, however, supposes that only wobble-effects are important. Occasionely, sudden "jumps"’could be observed, perhaps coincident with the desaturation of the reaction wheels. This non-periodic effect, of course, can not be corrected using the phase-method. It is not necessary to correct the image using the target counts themselves if there is a brighter source within the field of view. However, these cases are rare, and the brighter source is off-axis which, by itself, leads to a broadening of the PSF. A computer code (IDL), which allows the aspect correction of HRI images, is given in the annex.

**The HRI-PSF, latest stage**

To verify the above correction procedure, we selected from the HRI archive all known point-like sources with countrate larger than 1 ct/s. This led to a sample of 13 sources: 7 of them are stars, the rest are essentially QSO's. The application of the correction procedure to this sample allow us to derive an analytical formulation of the ROSAT-HRI PSF. This is found to have a width of the core of 4.8 arcsec (half energy width, HEW), about 20% better than the values measured prior launch. From this analysis we have derived a "perfect" PSF which represents the performance of the ROSAT telescope (HRI) which should be the reference for all HRI observations including those of point like sources, since also these show broad surface brightness profiles. We have compared the measured profiles also with the "standard" PSF (e.g., the one implemented in EXSAS). This PSF is based on the analysis done by David et al. (1999) but includes also the mirror scattering (which, for this exercise, is not important). The result are given in table 2, the plots are shown in figure 1 through figure 13. In all cases, our correction mechanism could improve the sharpness of the profiles, although only slightly sometimes. Two out of the 13 sources seem to be really extended since their profile remained broader than even David's et al. PSF.

Table 2: list of HRI-observations used for this analysis

| **observation** | **target** | **profile uncorrected data** | **profile corrected data** |
|---|---|---|---|
| rh200897 | Gliese 735 (star) | slightly broader than standard PSF | unresolved |
| rh201864 | HR 5144 (star) | broader than standard PSF | extended? |
| rh202057 | 47 CAS FOVn (binary) | broader than standard PSF | unresolved |
| rh202061 | AR Lac(G2IV) | slightly narrower than standard PSF | unresolved |
| rh300378 | AM Herculis (CV) | consistent with standard PSF | unresolved |
| rh300516 | CAL83 (SSS) | broader than standard PSF | unresolved |
| rh300542 | QS TEL (AM Her type) | broader than standard PSF | unresolved |
| rh400145 | G109.1-10 (star) | slightly narrower than standard PSF | unresolved |
| rh700234 | 3C273 (QSO) | narrower than standard PSF | unresolved |
| rh701661 | 3C273 (QSO) | broader than standard PSF | unresolved |
| rh701990 | 1H 0419-577 (Seyfert 1.5) | broader than standard PSF | extended? |
| rh702156 | PG1440+356 (QSO) | consistent with standard PSF | unresolved |
| rh702433 | RX309306+4950 (Blazar) | slightly narrower than standard PSF | unresolved |

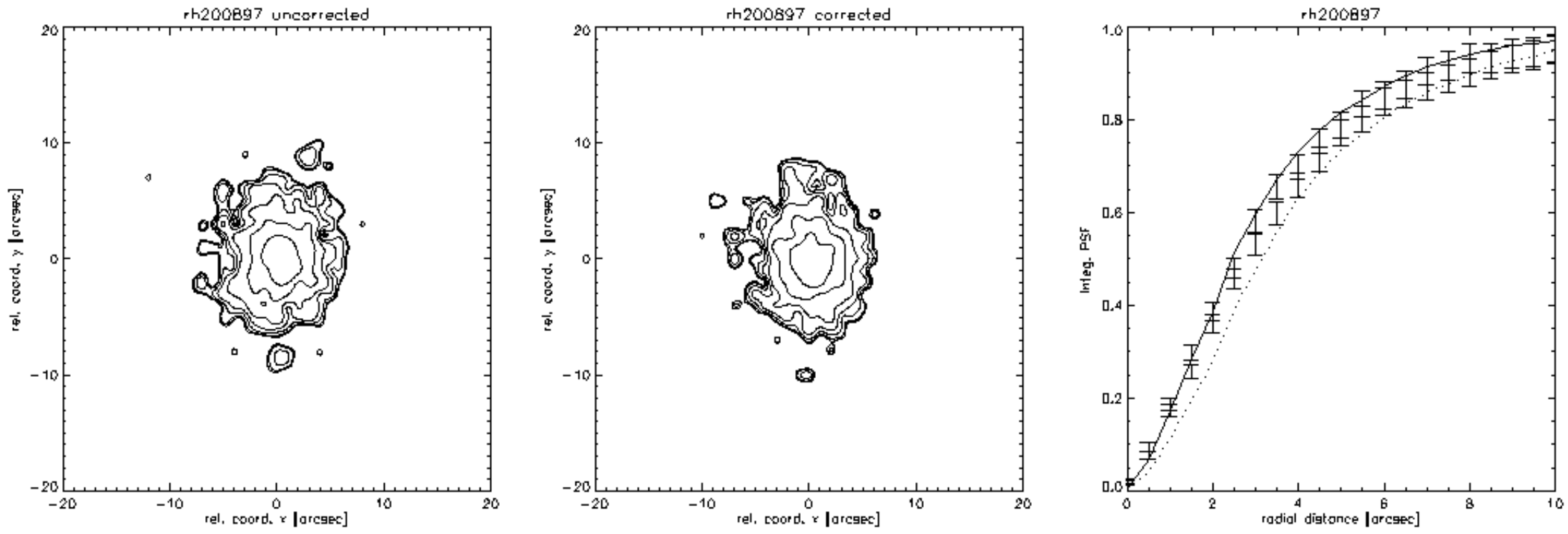

Figure 1: contour plots of the observation rh200897 before (left) and after the correction (center). The integrated point response function could be improved only slightly as indicated by the error bars. The solid line represents the "perfect" theoretical PSF, the dotted curve the standard PSF (see text).

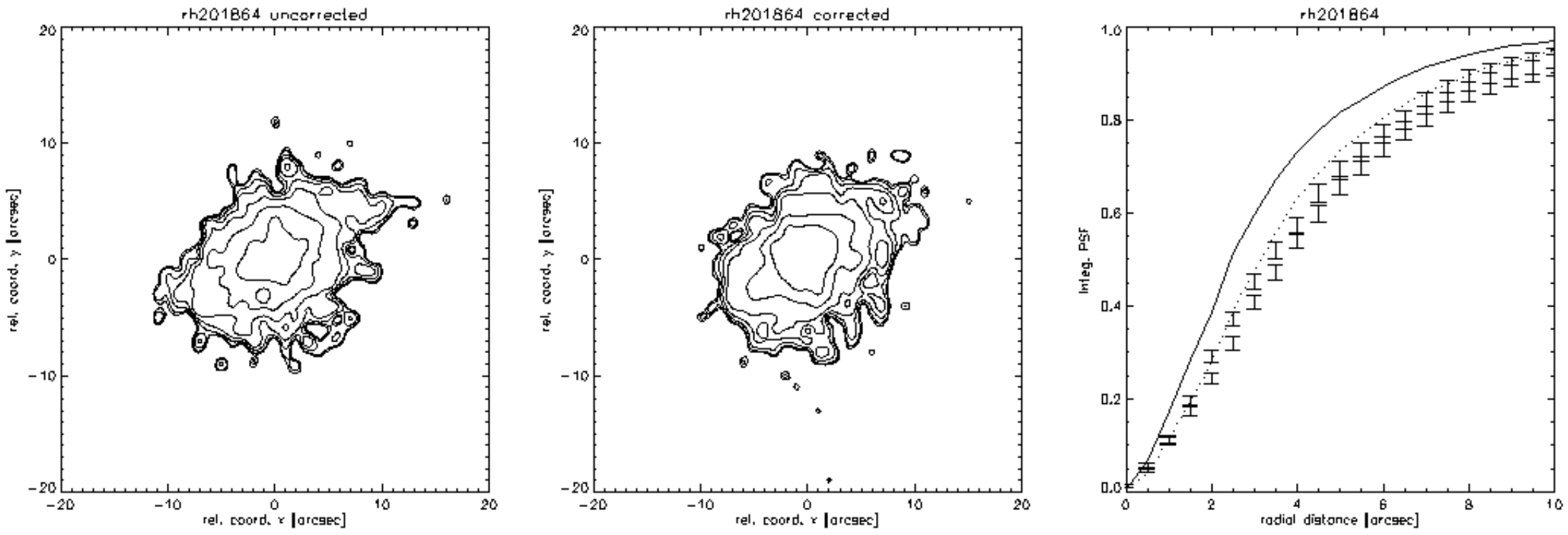

Figure 2: same as figure 1 for the the star HR 5144. Since the PSF is broader than theoretically predicted even after the correction, an extended source could be assumed which, however, seems to be a bit unlikely.

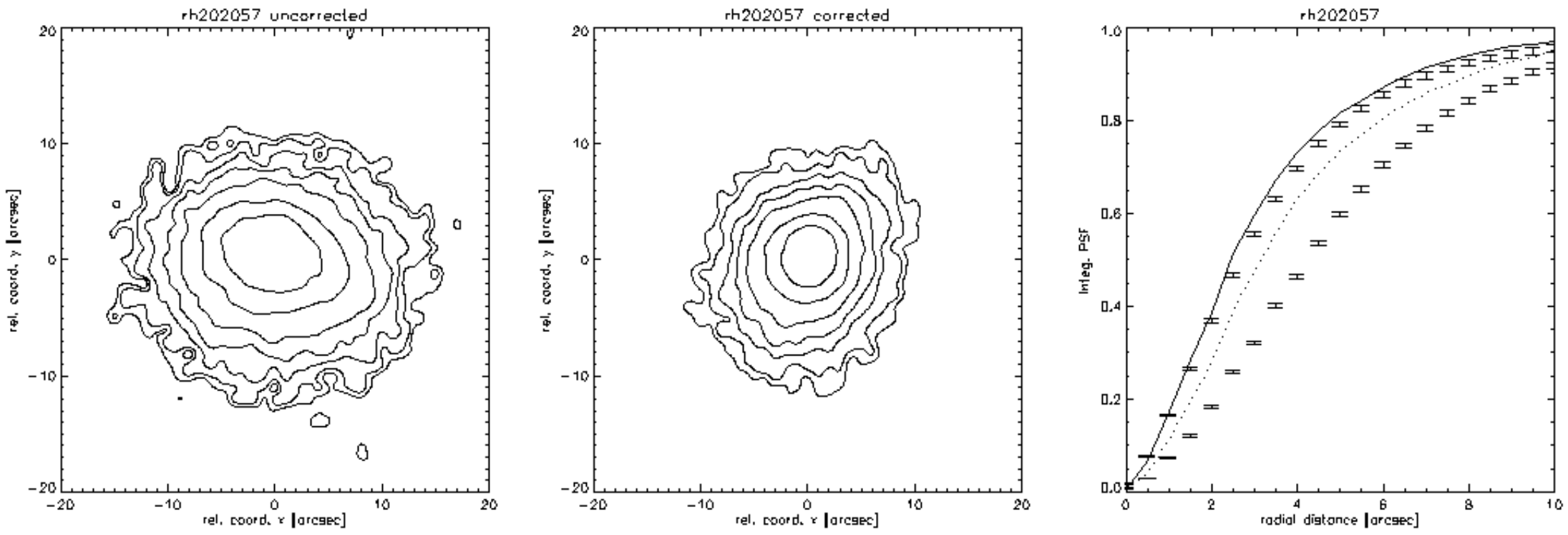

Figure 3: 47 Cas could be corrected absolutely perfectly!

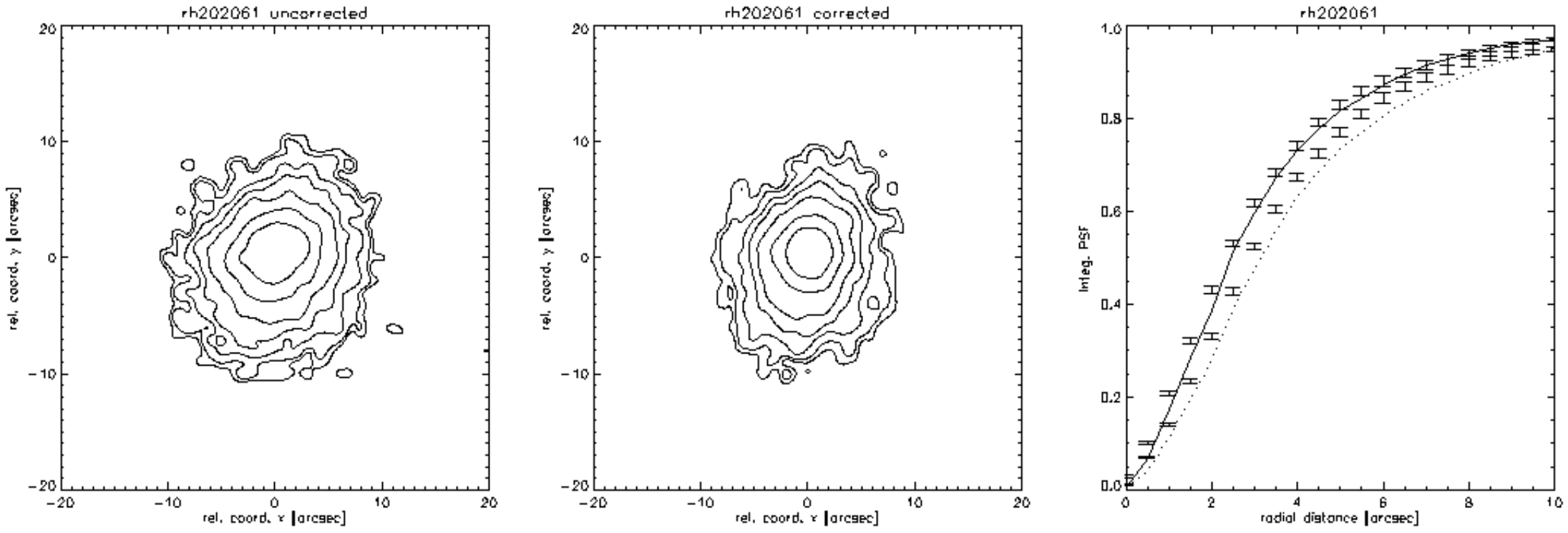

Figure 4: AR Lac shows a rather narrow PSF even before correction but could nevertheless a bit improved. The result is perfect.

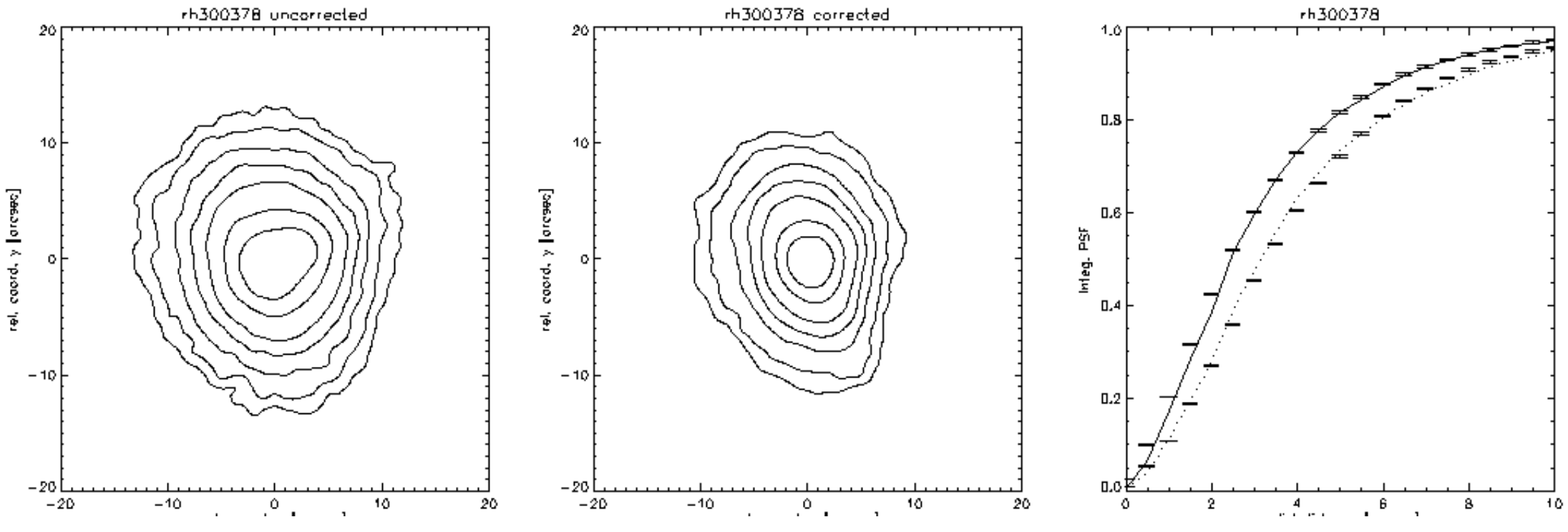

rel. coord. x [arcsec] rel. coord. x [arcsec] radial distance [arcsec]

Figure 5: also AM Her shows a reasonably narrow distribution, which is perfect after correction.

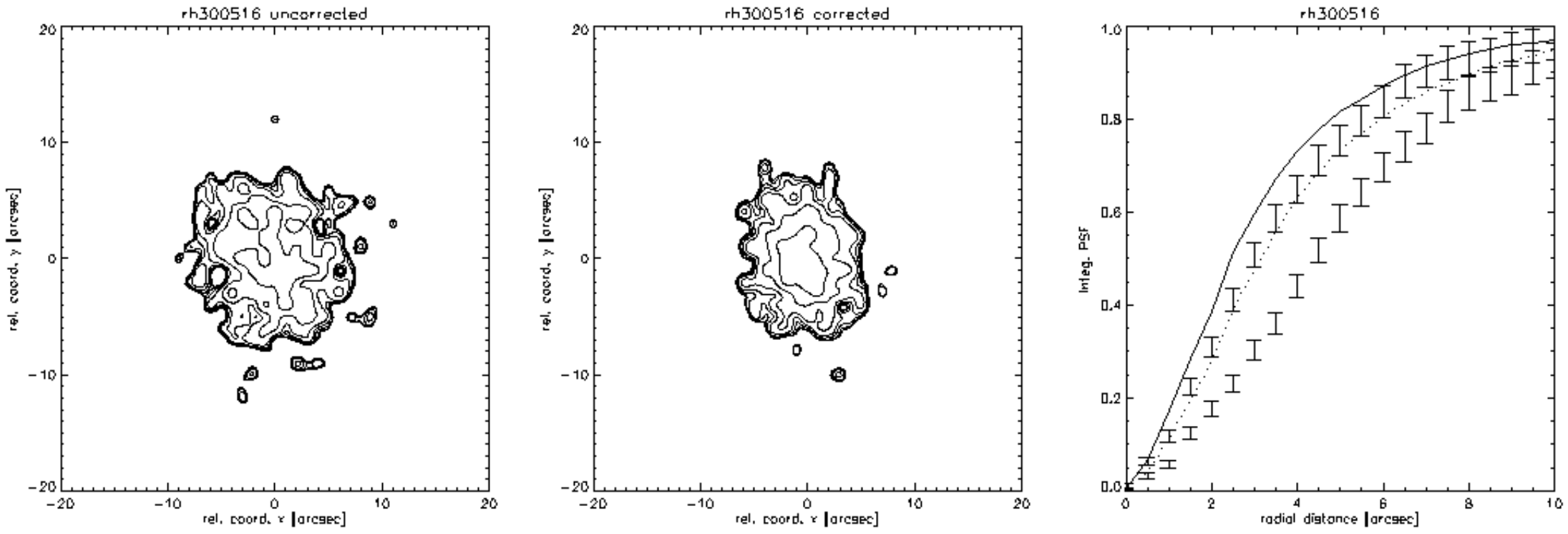

Figure 6: the original data of Cal 83 is a rather broad distribution, which could not be corrected perfectly.

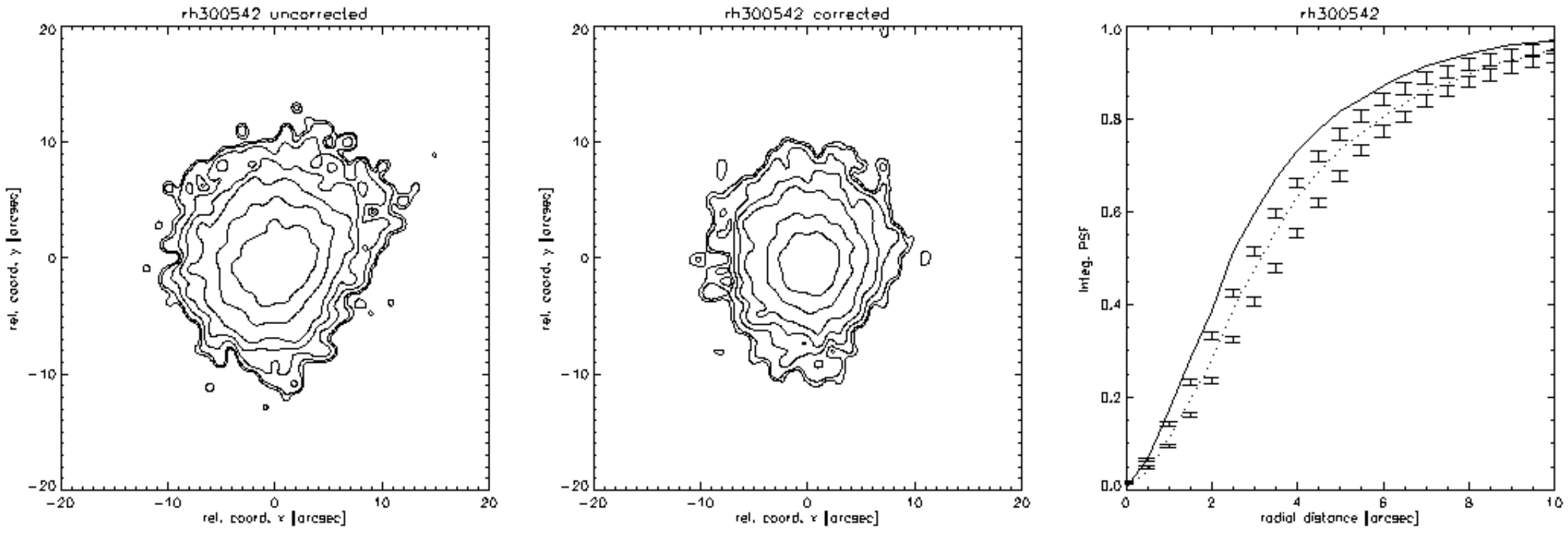

Figure 7: QS Tel has a broadened distribution before correction and an almost perfect PSF after.

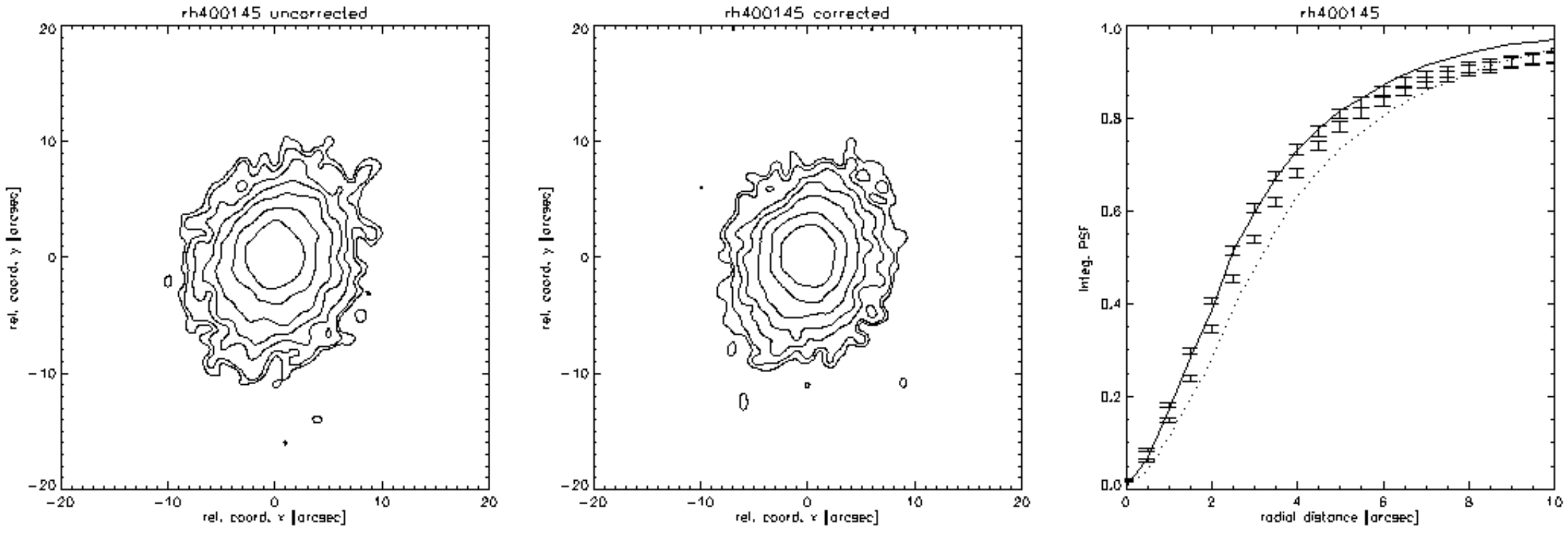

Figure 8: G109.1-10 is perfect – before and after the correction.

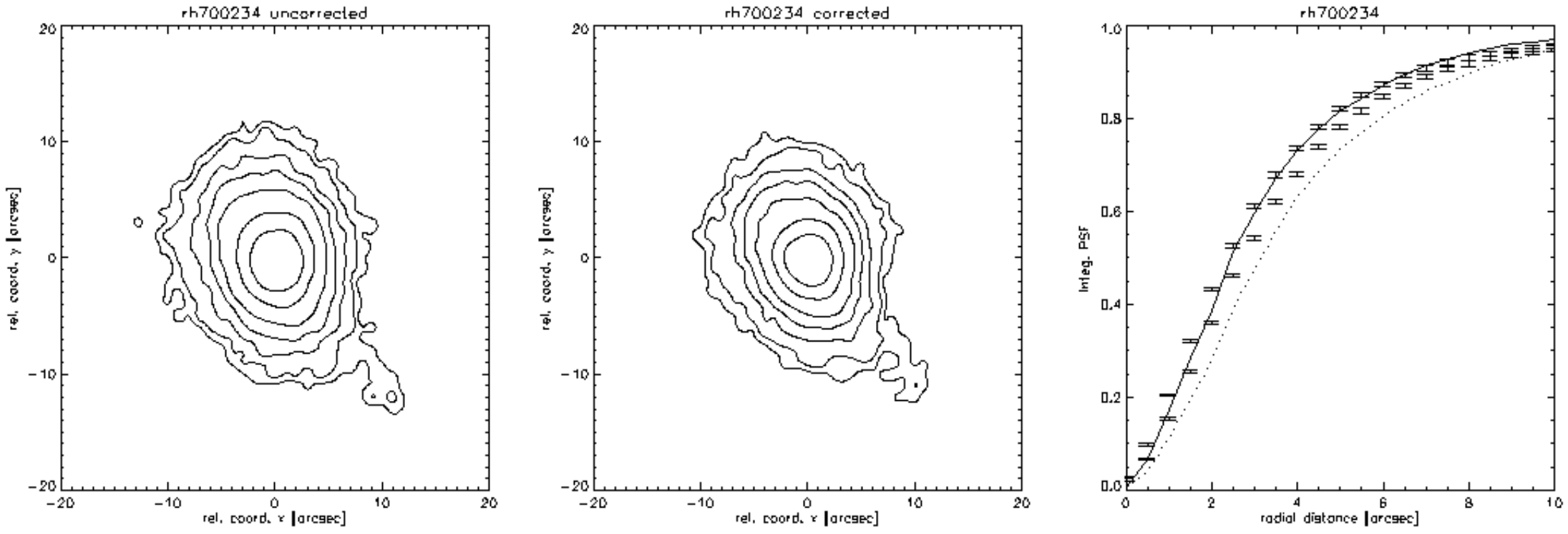

Figure 9: also this observation of 3C 273 was perfect and could almost not be improved.

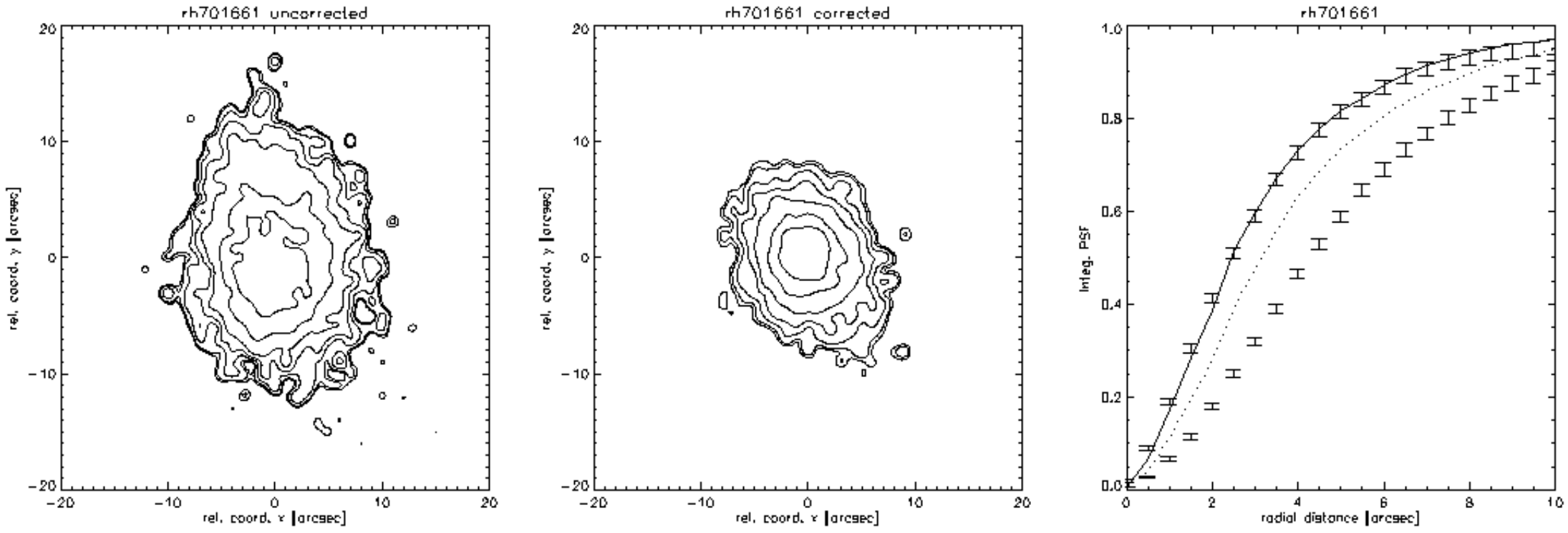

Figure 10: unlike in Figure 9, another observation of 3C 273 shows a very broad distribution , which, however, could be perfectly corrected.

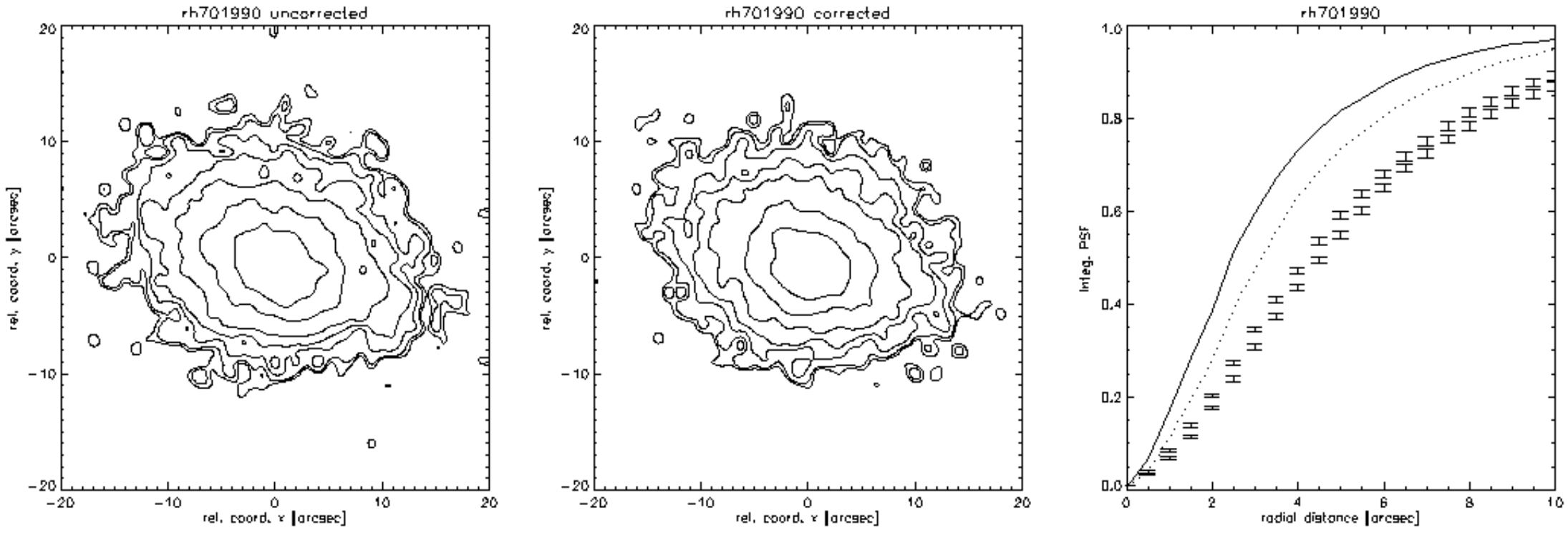

Figure 11: the Seyfert 1.5 galaxy 1H 0419-577 seems to be really extended.

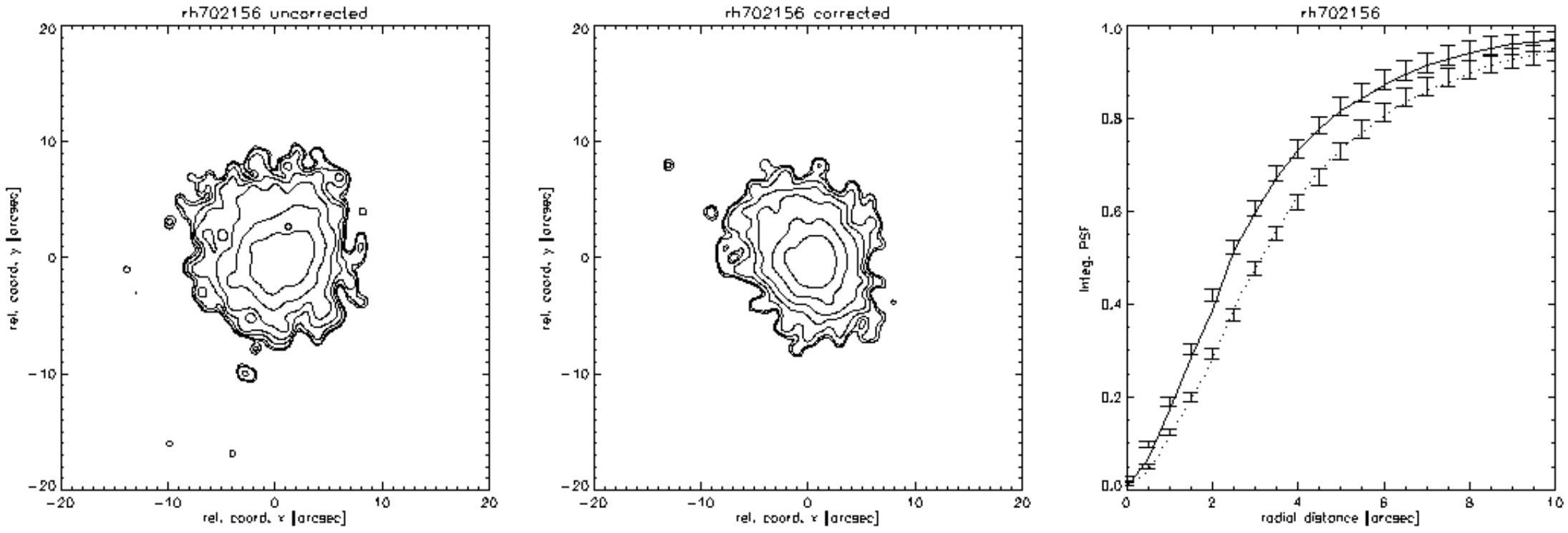

Figure 12: the Quasar PG1440+356 was slightly broadened but could also be perfectly corrected.

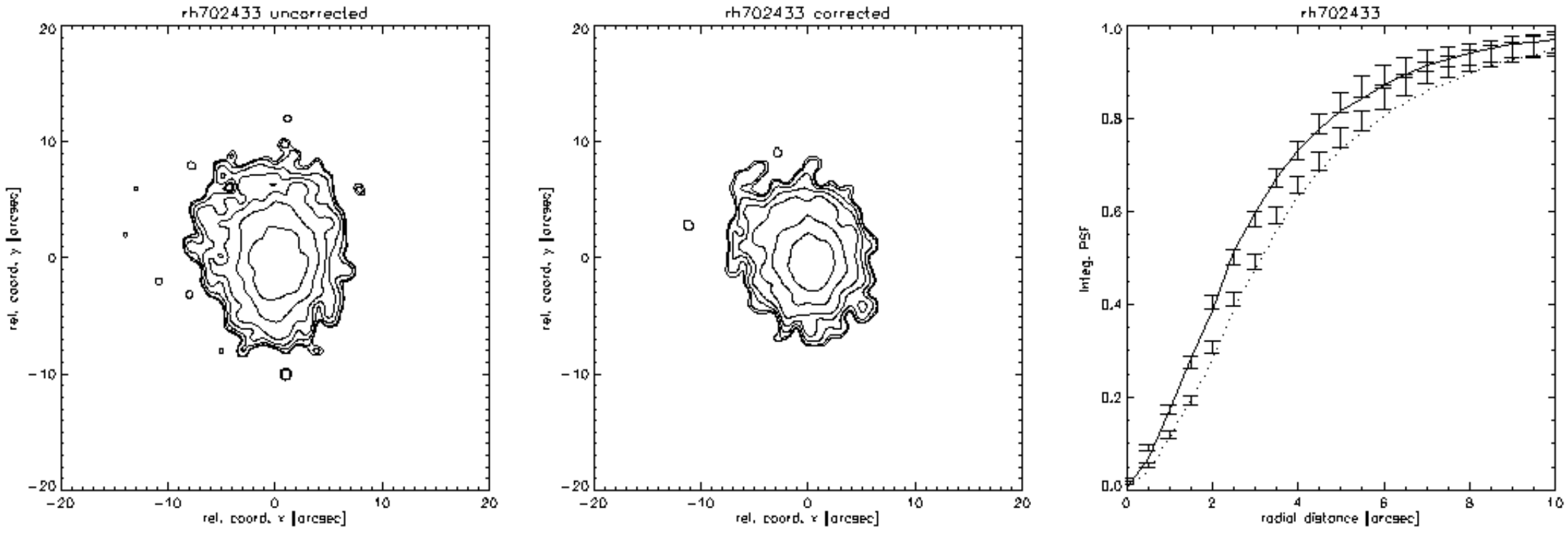

Figure 13: the observation blazar RX309306+4950 gave a reasonably narrow PSF.

After correction it became also perfect.

## Conclusion

From the analysis of the sources above indicated, we can draw the following conclusions:

1. After correction, all the sources are unresolved with the exception of the Seyfert Galaxy 1H0419-577. Apart from the two "extended" sources, the Seyfert 1.5 galaxy 1H 0419-577 and the star HR 5144, the width and shape of the corrected light profile distribution of the unresolved cases is in general midway between that derived from the perfect PSF and the standard PSF of David et al. (1999).
2. In four cases, the uncorrected HRI light profile shows slightly narrower than the standard PSF. In two additional cases, the uncorrected images are compatible with the standard PSF.All together, it may indicate that David's et al estimated PSF is slightly broader than it should be.
3. In all cases, the light profile distribution of the uncorrected data is wider than our derived PSF.

The success of the correction procedure in these sources has been possible because of the high photon statistics of the data. In sources with counts rate of less than 1, the correction should be done with caution as the number of photons per bin goes down dramatically. In such cases, attempts to re-center the image may lead to spurious features in the data. The examples above illustrate the kind of image broadening one may expect from HRI data without recentering of the arrival photons. Only in cases of high photon statistics the recovering of the intrinsic ROSAT HRI PSF is guaranteed.

In summary, the performance of the telescope is better than specified but slightly worse than it could be in view of the excellent properties of the hardware. For bright sources, residual aspect errors can be corrected, for dim sources one has to be cautious with statements about their possible extension on angular scales below 10 arcsec.

## References

1. A. Hefler et al.: Status der Verifikation des System Fehlerbudgets, Technical Note Dornier System, TN-2002-2160 DS/044 (1989)
2. David, L.P., et al.: The ROSAT High Resolution Imager (HRI), Calibration Report, U.S. ROSAT Data Center/SAO (revision June 1999)
3. J.A. Morse: A Method for Correcting Aspect Solution Errors in ROSAT HRI Observations of Compact Sources, PASP 106, 675 (2000)
4. D.E. Harris et al.: Spatial corrections of ROSAT HRI observations, Astron. Astrophys. Suppl. 133, 431 (1998)
5. G. Boese: The ROSAT Point Spread Function and Associates, Astron. Astrophys. 141, 507 (2000)

## Annex

In the following, various IDL computer-codes are given for the calculation of PSFs and the correction of HRI images.

### 1. Standard PSF

This is an IDL-representation of the PSF of David at al (1999), which is implemented in EXSAS.

```
*******************************************************************
pro hripsf,x,psf                                        ; array x: radial distance,
                                                          array psf: surface brightness at x
erg=1.                                                  ; photon energy set to 1 keV (fixed)
mirfrac=0.04*erg^1.43                                   ; fraction of mirror scattering
break1=39.95/erg& break2=861.9/erg                      ; transition points of broken power laws
alpha2=2.119+0.212*erg
a1=0.9638& a2=0.1798& a3=0.0009*10^0.2                  ; relative fraction of the three
                                                          contributions to the PSF
aux=1.+break2^2/break1^2
a4=mirfrac/(!pi*(alog(aux)+2./(aux*(alpha2-2.))))
sq1=2.1858^2& sq2=4.0419^2*10.^0.1
s3=31.69/10.^0.1
x1=0.5*x^2/sq1
x2=0.5*x^2/sq2
x3=x/s3
fold_norm=(1.-mirfrac)/((a1*sq1+a2*sq2+a3*s3^2)*2*!pi)
psf=(a1*exp(-x1)+a2*exp(-x2)+a3*exp(-x3))*fold_norm+a4/(break1^2+x^2)
return
end
```

### 2. Modified PSF

This perfect PSF corresponds mostly to the standard PSF, but with some modified parameters (marked in red).

```
*******************************************************************
pro hripsf_change,x,psf                                 ; array x: radial distance,
                                                          array psf: surface brightness at x
erg=1.0                                                 ; photon energy set to 1 keV (fixed)
mirfrac=0.04*erg^1.43                                   ; fraction of mirror scattering
break1=39.95/erg& break2=861.9/erg                      ; transition points of broken power laws
alpha2=2.119+0.212*erg
a1=0.9638& a2=0.1798& a3=0.0009                         ; relative fraction of the three
                                                          contributions to the PSF
aux=1.+break2^2/break1^2
a4=mirfrac/(!pi*(alog(aux)+2./(aux*(alpha2-2.))))
sq1=1.7^2& sq2=3.7^2
s3=31.69/10.^0.1
x1=0.5*x^2/sq1
x2=0.5*x^2/sq2
x3=x/s3
fold_norm=(1.-mirfrac)/((a1*sq1+a2*sq2+a3*s3^2)*2*!pi)
psf=(a1*exp(-x1)+a2*exp(-x2)+a3*exp(-x3))*fold_norm+a4/(break1^2+x^2)
return
end
```

### 3. Correcting an HRI image

This routine corrects an HRI image using the following steps:

1. The events are sorted in time (ascending)
2. The overall centroid of the source is calculated from the (uncorrected) event coordinates
3. For each sample of s consecutive events, their centroid is calculated. All events in that time-interval are corrected.
4. The same is done in phase intervals (of the wobble period) rather than time intervals. This corrects wobble-effects with better statistical significance than the previous dejittering.
5. The final centroid is calculated.

```
*******************************************************************
pro correct_attitude,x,y,t,cxall,cyall,s                  ; x,y: celestial coordinates of events,
                                                            t: eventtime
                                                          ; cxall, cyall: centroid of the
                                                                          (uncorrected) image
```

```
index=sort(t) & x=x(index) & y=y(index) & t=t(index)      ; sort events in time
centroid,x,y,cxall,cyall,50                               ; calculate centroid of all events
                                                            within 25"
print,' Preliminary centroid: ',cxall,cyall,format='(a,2f7.1)'
;---------------------------------------------------------- DE-JITTER
i1=where(sqrt((x-cxall)^2+(y-cyall)^2) lt 50)             ; only events within 25" around centroid
n=long(n_elements(x(i1))/s)                               ; number of complete samples
i2=lindgen(n*s)                                           ; events within last incomplete sample
                                                            are lost!
xo=x(i1(i2)) & yo=y(i1(i2)) & to=t(i1(i2))                ; only events within complete samples
xo=reform(xo,s,n) & yo=reform(yo,s,n) & to=reform(to,s,n) ; reform x,y,t into 2-dim arrays
cx=total(xo,1)/s & cy=total(yo,1)/s & ct=total(to,1)/s    ; calc. centroid for each sample
cx1=interpol(cx,ct,t) & cy1=interpol(cy,ct,t)             ; expansion to all events within OBI
x=float(x-cx1+cxall) & y=float(y-cy1+cyall)               ; correct x and y position of all events
                                                            in OBI
;---------------------------------------------------------- DE-WOBBLE
tp=t mod 402                                              ; wobble period [sec]
for j=0, 401, 40 do begin                                 ; ten phase bins
i1=where(tp ge j and tp le j+40,c)                        ; all events in current phase bin
i2=where(sqrt((x(i1)-cxall)^2+(y(i1)-cyall)^2) lt 50)     ; all events within circle around
                                                            centroid
cx=mean(x(i1(i2))) & cy=mean(y(i1(i2)))                   ; centroid of phase bin
x(i1)=x(i1)-cx+cxall                                      ; correct x position in phase bin of OBI
y(i1)=y(i1)-cy+cyall                                      ; correct y psoition in phase bin of OBI
endfor
;-----------------------------------------------------------
centroid,x,y,cxall,cyall,50                               ; final centroid of all events
;print,' Final centroid: ',cxall,cyall,format='(a,2f7.1)'
return
end
```

## 4. Centroiding

Calculates the centroid of an sample of events within a given circle.

```
********************************************************************
pro centroid,x,y,cx,cy,rad                                 ; x,y: celestial coordinates of events
                                                           ; cx, cy now: predefined (coarse)
                                                                                  centroid
index=where(sqrt((x-cx)^2+(y-cy)^2) lt 2*rad)              ; events within circle around coarse
                                                             position
cx=mean(double(x(index))) & cy=mean(double(y(index)))      ; preliminary centroid
index=where(sqrt((x-cx)^2+(y-cy)^2) lt rad)                ; events within circle around fine
                                                             position
cx=mean(double(x(index))) & cy=mean(double(y(index)))      ; cx, cy now: final centroid
return
end
```